\documentclass [12pt]{article}
\usepackage{graphicx,amssymb,amsfonts,latexsym,amsmath,amsthm,times}
\usepackage{epsfig}
\usepackage{fancyhdr}
\usepackage{color}
\newcommand{\new}[1]{#1}%\textcolor{blue}{#1}}
\newcommand{\newB}[1]{#1}%{\textcolor{blue}{#1}}
\usepackage[english]{babel}
\usepackage{booktabs}
\numberwithin{equation}{section}

\DeclareSymbolFont{extraup}{U}{zavm}{m}{n}
\DeclareMathSymbol{\vardiamond}{\mathalpha}{extraup}{87}

\newcommand{\klamm}[1]{\left(#1\right)}
\newcommand{\klammCurl}[1]{\left\{ #1 \right\}}
\newcommand{\klammSquare}[1]{\left[ #1 \right]}

\newcommand{\defi}{{\mathrel{\mathop:}=}}

\newcommand{\nDensity}{n}

\newcommand{\chemPot}{\mu}

\newcommand{\depthLJ}{\varepsilon}
\newcommand{\depthLJW}{\depthLJ_{\text{w}}}
\newcommand{\LJdiam}{\sigma}

\newcommand{\surfaceTension}{\gamma}

\newcommand{\pos}{{\bf r}}

\newcommand{\DisjoiningPressure}{\Pi}
\newcommand{\chemPotN}{\chemPot}

\newcommand{\nDensityV}{\nDensity_{\text{vap}}}
\newcommand{\nDensityL}{\nDensity_{\text{liq}}}
\newcommand{\nDensityW}{\nDensity_{\text{w}}}

\newcommand{\surfaceTensionLV}{\surfaceTension_{\text{lv}}}
\newcommand{\surfaceTensionWV}{\surfaceTension_{\text{wv}}}
\newcommand{\surfaceTensionWL}{\surfaceTension_{\text{wl}}}

\newcommand{\filmThickness}{\ell}
\newcommand{\GrandPotential}{\Omega}

\newcommand{\FE}{\mathcal{F}}
\newcommand{\FEhs}{\mathcal{F}_{\text{HS}}}
\newcommand{\FEattr}{\mathcal{F}_{\text{attr}}}

\newcommand{\Vext}{V_{\text{ext}}}

\newcommand{\BHattr}{\phi_{\text{attr}}}

\newcommand{\BHattrWF}{\phi_{\text{attr}}^{\text{wf}} }

\newcommand{\dI}{\text{d}}			
\newcommand{\thYoung}{\theta_{\text{Y}}}

\newcommand\etal{\mbox{\textit{et al.}}}

\usepackage{subfig}
\usepackage{caption}

\begin{document}

\thispagestyle{plain}

%*******************************************************************
%TITLE OF THE ARTICLE
%*******************************************************************
\vspace*{2cm} \normalsize \centerline{\Large \bf
Nanoscale fluid structure of liquid-solid-vapour}
%Large-contact angle fluid structure}
\centerline{\Large \bf contact lines for a wide range of contact angles}

\vspace*{1cm}
%*******************************************************************
%AUTHORS - THE CORRESPONDING AUTHOR NEEDS TO SPECIFY HIS/HER E-MAIL ADDRESS AS A FOOTNOTE
%*******************************************************************

\centerline{\bf A. Nold$^a$, D. N. Sibley $^a$, B. D. Goddard $^b$ and S. Kalliadasis $^a$
\footnote{Corresponding author. E-mail: s.kalliadasis@imperial.ac.uk}}

\vspace*{0.5cm}

%*******************************************************************
%ADDRESS OF THE AUTHORS
%*******************************************************************
\centerline{$^a$ Department of Chemical Engineering, Imperial College London, London SW7 2AZ, UK}
\centerline{$^b$ The School of Mathematics and Maxwell Institute for Mathematical Sciences}
\centerline{The University of Edinburgh, Edinburgh EH9 3JZ, UK}

%*******************************************************************
%ABSTRACT
%*******************************************************************

\vspace*{1cm}

\noindent {\bf Abstract.} {We study the nanoscale behaviour of the density of
a simple fluid in the vicinity of an equilibrium contact line for a wide
range of Young contact angles $\thYoung  \in [40^\circ,135^\circ]$. Cuts of
the density profile at various positions along the contact line are
presented, unravelling the apparent step-wise increase of the film height
profile observed in contour plots of the density. The density profile is
employed to compute the normal pressure acting on the substrate along the
contact line. We observe that for the full range of contact angles, the
maximal normal pressure cannot solely be predicted by the curvature of the
adsorption film height, but is instead softened --
\newB{likely} by the width of the liquid-vapour interface. Somewhat
surprisingly however, the adsorption film height profile can be predicted to
a very good accuracy by the Derjaguin-Frumkin disjoining pressure obtained
from planar computations, as was first shown in [Nold~\etal, Phys. Fluids,
{\bf 26}, 072001, 2014] for contact angles $\thYoung < 90^\circ$, a result
which here we show to be valid for the full range of contact angles. This
suggests that while two-dimensional effects cannot be neglected for the
computation of the normal pressure distribution along the substrate,
one-dimensional planar computations of the Derjaguin-Frumkin disjoining
pressure are sufficient to accurately predict the adsorption height profile.}
\vspace*{0.5cm}
%*******************************************************************
%KEYWORDS
%*******************************************************************
\noindent {\bf Key words:}{adsorption, contact line, simple fluid, disjoining pressure, Derjaguin-Frumkin, Hamiltonian}

% these are examples, put your key words

% these are examples, put your numbers

%********************************
%*********OWN VERSION*************
%********************************

%*******************************************************************
%DO NOT FORGET TO RESET THE EQUATION COUNTER TO 0 AT THE HEAD OF EACH SECTION
%*******************************************************************

\section{Introduction}
\setcounter{equation}{0}

Consider a fluid interface in contact with a solid substrate. This scenario
describes a container filled with liquid, a drop sitting on a leaf, or a
vapour bubble \new{inside a liquid filled bottle.} Imagine observing a point
in the vapour phase. As the liquid phase is approached, a \new{rapid, yet}
smooth transition in the density occurs at the liquid-vapour interface.
Staying \new{on} this interface and approaching the substrate, would reveal a
variety of physical effects that \new{become significant.} First, the fluid
feels an attractive force of the wall particles. At the same time, the nature
of the solid substrate forces the fluid particles to \new{`jam'} and restrict
their mobility as the wall is approached.

In this work, we are interested in the effect \newB{of} the wall attractive
forces on the density profile in the vicinity of a three-phase contact line
for a wide range of contact angles. Developing a fundamental understanding of
these small-scale phenomena at equilibrium is important to predict the
dynamic nanoscale behaviour of the moving contact line, which is still a
controversial problem with a wide range of physical explanations being
offered (for a review, see Bonn~\etal~\cite{Bonn.20090527} or Snoeijer and
Andreotti~\cite{SnoeijerAndreotti:2013}). In this context, our intention is
twofold: First, to illustrate and give a general understanding for the
density structure of the fluid as well as its
\new{form} and scale of variations in the vicinity of the contact line;
\new{and second}, to illustrate the impact of the contact line on the normal
pressure distribution acting on the substrate. The latter point is directly
connected with the definiton of the disjoining pressure. The uniqueness of
disjoining pressure definitions was recently discussed critically in several
papers~\cite{Herring:2010vn,Henderson:2011:EPJST:DisjoiningPressure,MacDowell:2011:ResponseEPJST,Henderson:EPJST:ResponseMacDowell,Henderson:NoteContinuingContactLine,Nold:2014:FluidStructure}.

To describe the interaction between a solid substrate and a fluid interface,
we choose to model a simple fluid, \new{i.e.\ }a system of identical
particles in contact with a homogeneous, perfectly flat\new{,} hard wall. The
particles of the fluid are modelled as hard spheres interacting with a
Lennard-Jones type potential decaying with $r^{-6}$, where $r$ is the
interparticle distance.
\newB{The wall and fluid particles are assumed to interact via a similar
Lennard-Jones type potential.}

Contact line models, including nonlocal contributions to the free energy beyond those of the disjoining pressure, have previously been studied
analytically~\cite{Merchant:1992kx,Pismen:2001fk,Snoeijer:2008fk,Getta:1998ly,SnoeijerAndreotti:2013}. However,
for the sake of analytical attainability, only simple models of the free energy model are considered and 
restrictive assumptions on the nature of the density profile at the contact line are made. 

In contrast, we consider the density structure at the contact line numerically employing
classical density functional theory (DFT), an approach derived from
the statistical mechanics of fluids~\cite{Evans}. DFT has
proven to be a numerically efficient way to model equilibrium properties of
inhomogeneous fluid systems. It can be viewed as middle ground between
continuum hydrodynamics, which is inapplicable at small fluid volumes,
and particle-based Monte-Carlo (MC) or Molecular Dynamics (MD) simulations,
which despite dramatic improvements in
computational power are still restricted to small fluid volumes.
In fact, compared to MC or MD simulations, for which the numerical complexity scales
with the number of particles modelled, DFT gives the ability to solve
directly for the density distribution,
with the advantage that its computational complexity is formally independent of the number of particles.
Thus, modelling larger systems, such as contact lines, becomes feasible.

The predictive qualities of the DFT results depend on the accuracy of the
free-energy model employed. Here, we model the hard-sphere free energy with a
fundamental measure theory (FMT)~\cite{Rosenfeld:1989qc}, while the
attractive forces are included as a Barker-Henderson
perturbation~\cite{Barker:1967rq} in a mean-field manner. DFT-FMT has been
applied successfully in studies of critical point wedge
filling~\cite{Malijevsky:2013:CriticalPointWedgeFilling}, phase transitions
in nanocapillaries~\cite{PeterPRE}, thin films on planar
substrates~\cite{Peter2012} and density computations in the vicinity of
liquid wedges~\cite{Merath:2008}. A previous study by Pereira~\etal~\cite{Antonio2010} on
equilibrium contact lines utilised a DFT local-density approximation (LDA)
which is not appropriate to describe structuring in the fluid and fails to
describe the oscillatory behavior of the density in the immediate vicinity of
a wall.

The present work parallels our previous study
in~\cite{Nold:2014:FluidStructure} where DFT-FMT was used to analyse the
fluid structure in the immediate vicinity of a contact line for $\thYoung <
90^\circ$. Here we investigate a wide spectrum of contact angles $40^\circ <
\thYoung < 135^\circ$ and we shed further light on the density structure in
the vicinity of the contact line and its dependency \new{on} the wall
strength. A discussion of the special case of a $90^\circ$ contact angle is also included.
We present density profiles slice by slice as we sweep through the
contact line region and we contrast the profiles with that of a planar liquid
film on a substrate with the same film thickness, but at an off-saturation chemical potential.
Interestingly, the two are not that different, which
suggests that results of the planar film case may be transferable to the contact line.
In particular, as
in~\cite{Nold:2014:FluidStructure} we shall scrutinize the ability of
Derjaguin-Frumkin theory~\cite{Derjaguin:1987:50YearsOfSurfaFceScience} for
planar liquid films on a substrate to predict the height profile at the
contact line. We offer a unified Derjaguin-Frumkin treatment of the contact
line for $\thYoung < 90^\circ$ and $\thYoung > 90^\circ$ by appropriately
extending the boundary conditions for the disjoining pressure equation to
account for the case $\thYoung > 90^\circ$. We further study the connection
between the Derjaguin-Frumkin disjoining pressure and the normal pressure
distribution acting on the substrate for non-planar liquid films, such as
given by the contact line, for $40^\circ < \thYoung < 135^\circ$.

In section 2 we give an overview of the DFT model employed to solve for the
equilibrium density profile. The numerical scheme to compute the contact
angles is introduced in section 3. A description of the density structure in
the vicinity of the contact line is given in section 4, before discussing
coarse-grained Hamiltonian approaches in section 5. Finally, a general
discussion of the results and concluding remarks are in section 6.

\section{Statistical mechanics framework \label{sec:DFTmodel}}
\vspace*{0.5cm} \setcounter{equation}{0}

As done for contact angles less than $90^\circ$ in
Ref.~\cite{Nold:2014:FluidStructure}, we employ classical DFT to investigate
the density distribution in the vicinity of an equilibrium contact line at
contact angles both greater and less than $90^\circ$. It is based on a
statistical mechanics description and has been successfully applied in the
study of inhomogeneous fluids. It is based on the theorem of Mermin
\cite{Mermin:1965lo}, which allows the Helmholtz free energy $\FE$ to be
written as a unique functional of the number density $\nDensity({\bf
r})$~\cite{Wu-DFT}. The equilibrium density distribution minimizes the grand
potential \cite{Evans}
\begin{align}
\GrandPotential[\nDensity] = \FE[\nDensity] + \int \nDensity(\pos) \klammCurl{\Vext({\bf r}) - \chemPotN} \dI\pos,  \label{eq:GrandPotential}
\end{align}
where $\chemPotN$ is the chemical potential and $\Vext$ is the external
potential, dependent on the position vector $\pos$. We then minimize
Eq.~(\ref{eq:GrandPotential}) by solving the Euler-Lagrange equation
\begin{align}
\frac{\delta \Omega[\nDensity]}{\delta \nDensity({\pos})} = 0,\label{eq:EulerLagrangeEquation}
\end{align}
where for a simple fluid of particles interacting with a Lennard-Jones
potential, the free energy is usually separated into a repulsive hard-sphere
part and an attractive contribution
\begin{align}
\FE[\nDensity] = \FEhs[\nDensity] + \FEattr[\nDensity].
\end{align}
To accurately model both the structure and thermodynamics of hard-sphere
fluids, we use the Rosenfeld FMT approach~\cite{Rosenfeld:1989qc} for the
hard-sphere contribution~\cite{Roth:2010fk}. The attractive interactions are
\new{modelled} with a mean-field Barker-Henderson approach
\cite{Barker:1967rq}
\begin{align}
\FEattr[\nDensity] &= \frac{1}{2 } \iint  \phi_{\text{attr}}({|\pos - \pos'|}) \nDensity(\pos)\nDensity(\pos') \dI\pos' \dI\pos, \label{eq:FEattr}
\end{align}
where the attractive interaction potential is given by
\begin{align}
\BHattr\klamm{r} =
\depthLJ \left\{ \begin{array}{ll}
0 & \text{for } r \leq  \LJdiam\\
4 \klamm{
\klamm{\frac{\LJdiam}{r}}^{12} -
\klamm{\frac{\LJdiam}{r}}^{6}
} & \text{for } r > \LJdiam
\end{array}  \right. .\label{eq:pattr}
\end{align}
Here, \new{$\depthLJ$ is the depth of the Lennard-Jones potential, $\LJdiam$ is the distance from the center of the particle at which the
Lennard-Jones potential is zero, and $r$ is a (scalar) radial distance}. The
simple fluid described by (\ref{eq:GrandPotential})--(\ref{eq:pattr}) has a critical point at \newB{$k_B T_c/\depthLJ = 1.0$}, where $k_B$ is the Boltzmann constant. \new{Computations} in
this work are performed at $T = 0.75 T_c$, at which the liquid and vapour
number densities are well-separated ($\nDensityL \LJdiam^3= 0.622$,
$\nDensityV \LJdiam^3= 0.003$) and at which the liquid-vapour surface tension \newB{resulting from planar DFT computations} is $\surfaceTensionLV = 0.3463 \depthLJ/\LJdiam^2$. All two-dimensional (2D) computations are
performed at the saturation chemical potential, at which the bulk \new{vapour} and
bulk liquid are equally stable.

The wall-fluid particle interaction is modelled
analogously to the fluid-fluid interaction as
\begin{align}
\BHattrWF\klamm{r} =
\depthLJW \left\{ \begin{array}{ll}
\infty & \text{for } r \leq  \LJdiam\\
4 \klamm{
\klamm{\frac{\LJdiam}{r}}^{12} -
\klamm{\frac{\LJdiam}{r}}^{6}
} & \text{for } r > \LJdiam
\end{array}  \right. , \label{eq:WallFluidInteraction}
\end{align}
where $\depthLJW$ is the depth of the wall-fluid interactions. Let us take a
Cartesian coordinate system with the $x$-$z$ plane parallel to the wall and the $y$-coordinate \new{in the} direction normal to
the wall. The external potential can then be obtained analytically from the
integration of the interactions over the uniform density distribution of wall
particles $\nDensityW$ for $y \leq -\LJdiam$, giving
\begin{align}
\Vext\klamm{y}
&=
\left\{
\begin{array}{ll}
\infty & y \leq 0\\
\frac{2}{3}\pi \alpha_{\text{w}} \LJdiam^3 \klammSquare{\frac{2}{15} \klamm{\frac{\LJdiam}{y + \LJdiam} }^9 - \klamm{\frac{\LJdiam }{y+ \LJdiam}}^3} & y > 0
\end{array}
\right.,
\end{align}
where $\alpha_{\text{w}} = \nDensityW \depthLJW$ is the strength of the wall
potential.
\begin{figure}[ht]
	\centering
		\includegraphics[width=9cm]{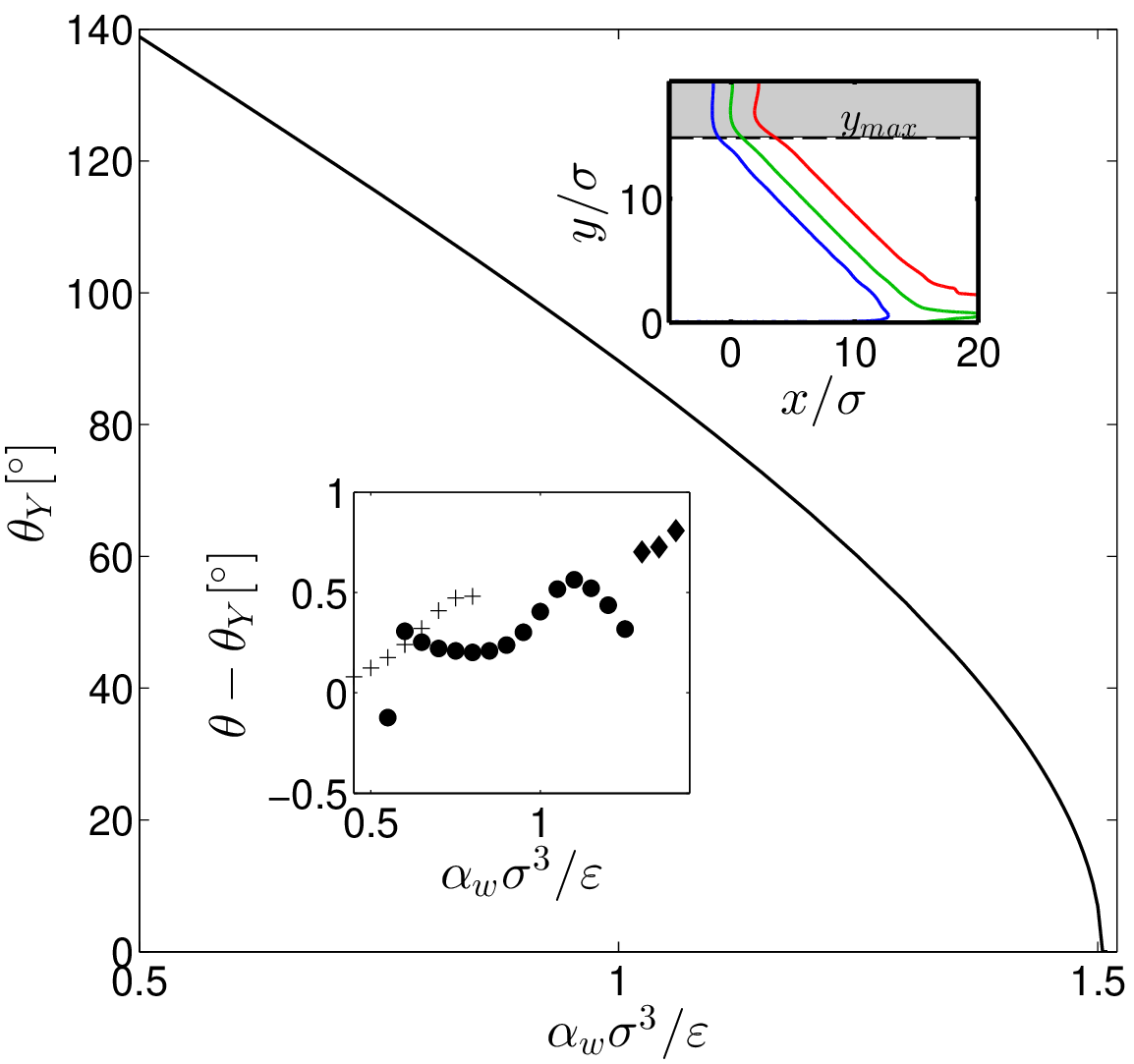}
		\caption{
		Plot of the Young contact angle $\theta_{\text{Y}}$ dependence on the strength of the wall attraction $\alpha_{\text{w}}$.
		Computations for \new{$\surfaceTensionLV$, $\surfaceTensionWV$ and $\surfaceTensionWL$} are done in a planar geometry, which are then inserted in (\ref{YoungEquation}).
    In the bottom left inset $\theta_{\text{Y}}$ is compared \new{to} 2D contact angle measurements $\theta$
		which are solved for $y < y_{\text{max}}= 15 \LJdiam$ and with $\theta_{\text{n}} = 90^\circ$ ($\bullet)$, $\theta_{\text{n}} = 120^\circ$ ($+$) and $\theta_{\text{n}} = 40^\circ$ ($\blacklozenge$).	
		The top right inset depicts the contour lines of the density profile for a Cartesian grid ($\theta_{\text{n}} = 90^\circ$) and $\alpha_{\text{w}} \sigma^3 /\depthLJ = 0.55$ \new{(giving $\theta_{\text{Y}}=134.2^\circ$)} for 		$y_{\text{max}}=15 \sigma$.		}	
	\label{fig:ContactAngleMeasurements90}	
\end{figure}

\section{Numerical Method \label{sec:NumericalMethod}}

To solve (\ref{eq:EulerLagrangeEquation}) numerically in a 2D domain, we
employ a spectral collocation method~\cite{Trefethen_2000}. We have used this
method successfully in our previous studies with both DFT-LDA and DFT-FMT
(e.g.~\cite{PeterPRE,Peter2012,Nold:2014:FluidStructure}). It should be
emphasized that because the equations we wish to solve are non-local, the
resulting matrices following discretization are dense, however the advantage
of the spectral collocation method is that through a convenient choice of
collocation points their number may be kept relatively low, leading to
significant reduction in the size of the matrices. The reduction in the
number of points becomes increasingly important when going to higher
dimensions (as the number of points in a product grid scales exponentially
with the dimension).

Consider the tensor product of two one-dimensional (1D) Chebychev grids on
the box $(\xi,\eta) \in [-1,1]\times [-1,1]$. This computational domain is
mapped onto the half space $[-\infty,\infty]\times[0,\infty]$ by
\begin{align}
x' =  L_1 \frac{\xi}{\sqrt{1-\xi^2}}
,\qquad
y' = L_2 \frac{1+\eta}{1-\eta}. \label{eq:CartesianMap}
\end{align}
Here, $L_1$ and $L_2$ are numerical parameters determining the spatial resolution of the collocation points close to $x'=0$ and in the vicinity of the wall, respectively.
This Cartesian grid in the physical half-space is then skewed by an angle $\theta_n$ using the map
\begin{align}
x = \frac{x'}{\sin \theta_n} + y' \cot \theta_n
, \qquad
y = y'. \label{eq:SkewedMap}
\end{align} The skewed grid allows us to have more discretization points near the
fluid-fluid and fluid-solid interface where higher density gradients are
expected. In our computations, we assume that the liquid-vapour interface is
at an angle of $\theta_n$ for values $y \geq y_{\text{max}}$, and only solve
for collocation points located at $y < y_{\text{max}}$, such that the resulting density profiles may only be interpreted for $y < y_{\text{max}}$.
In order to minimize the numerical inaccuracy caused by this cut-off, we iteratively adapt $\theta_n$ and increase 
$y_{\text{max}}$ to obtain a final result which is fully physically interpretable.

Physically, the contact angle of a liquid wedge is uniquely defined through
the surface tensions of the liquid-\new{vapour} phase\new{,
$\surfaceTensionLV$,} and the wall-fluid pair \new{($\surfaceTensionWV$ and
$\surfaceTensionWL$ being wall-vapour and wall-liquid surface tensions,
respectively)}, given by
\new{the Young} equation
\begin{align}
\surfaceTensionLV \cos \theta_{\text{Y}} = \surfaceTensionWV - \surfaceTensionWL, \label{YoungEquation}
\end{align}
where the surface tensions are quantities that can be extracted from
planar/\new{(1D)} DFT computations and $\theta_{\text{Y}}$ is defined as the
Young contact angle. Given that we restrict our attention to systems at
temperature $T/T_c = 0.75$, the only parameter on which
\new{$\theta_{\text{Y}}$} depends is the strength of the wall attraction
$\alpha_{\text{w}}$. In figure \ref{fig:ContactAngleMeasurements90}, we plot
the dependence of
\new{$\theta_{\text{Y}}$} on the wall attraction. As expected
intuitively, the contact angle decreases with increasing wall-fluid
attraction and reaches complete wetting at the critical value of
$\alpha_{\text{w,crit}} \sigma^3/\depthLJ = 1.50$. In 2D computations, the
contact angle of the liquid-vapour interface has to converge to
\new{$\theta_{\text{Y}}$} at large distances from the wall.

To check this, we have performed computations on a Cartesian grid, employing
(\ref{eq:CartesianMap}) \newB{and (\ref{eq:SkewedMap}) with $\theta_n =
90^\circ$}, and assuming that above a limiting value $y_{\text{max}}$, the
density at the collocation points corresponds to an equilibrium liquid-vapour
interface with a $90^\circ$ contact angle. The result of the density profile
for such a computation is depicted in the top right inset of figure
\ref{fig:ContactAngleMeasurements90}. By measuring the slope of the
isodensity line for $\nDensity = (\nDensityL + \nDensityV)/2$ in the interval
$y \in [10\sigma,14\sigma]$, we obtain an estimate for the contact angle in a
2D setting. The deviations to the
\new{$\theta_{\text{Y}}$} are shown in the bottom left inset of figure
\ref{fig:ContactAngleMeasurements90}, showing very good agreement.

\begin{figure}[ht]
	\centering
		\includegraphics[width=9cm]{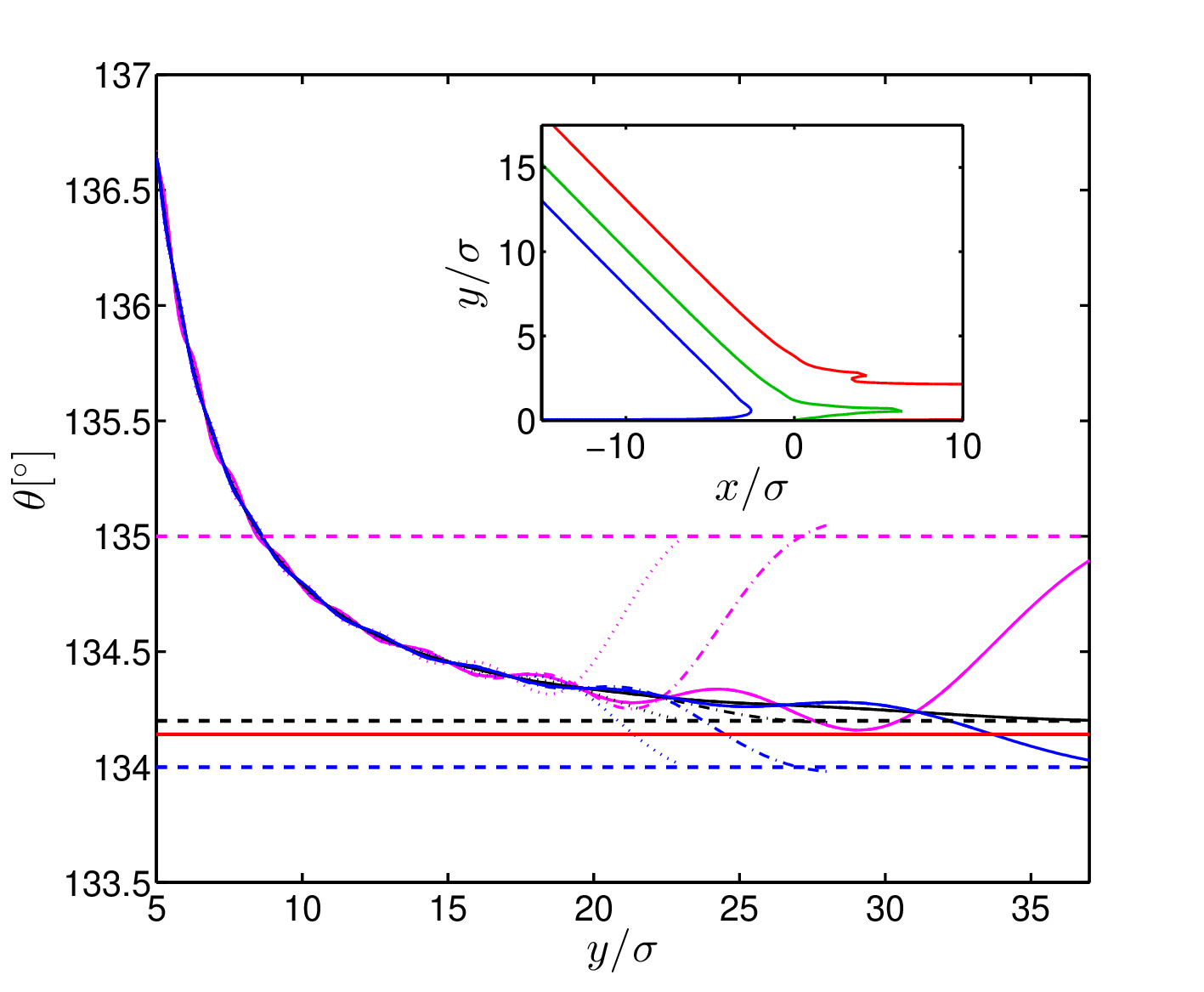}
		\caption{Slope of the isodensity line for $\nDensity = \klamm{\nDensityV + \nDensityL}/2$ for $y_{\max} = \{20,25,35\}$,
        represented by the dotted, dash-dotted and solid lines, respectively.
        Computations are done on a skewed grid with $\theta_n = 135^\circ$,
        $134^\circ$ and $134.2^\circ$, represented by horizontal dashed lines, and results for which are 
				drawn with magenta, blue and black lines,
        respectively. The substrate strength is
        $\alpha_w \sigma^3/\depthLJ = 0.55$, leading to \new{$\theta_{\text{Y}} =
        134.14^\circ$}, depicted by the red horizontal line. The inset
        shows a typical contour plot for the density,
        where the contour lines correspond to number densities $\klamm{\nDensity -
        \nDensityV}/\klamm{\nDensityL - \nDensityV} = \{0.05,0.5,0.95\}$ from left to
        right, respectively.}
	\label{fig:CA}	
\end{figure}

We have also performed computations on skewed grids, to increase the number
of collocation points in the vicinity of the contact line and the
liquid-vapour interface, by assuming that the liquid-vapour interface is at
an angle of $\theta_n$ for values $y \geq y_{\text{max}}$. This allowed us to
increase the value of $y_{\text{max}}$ to higher values. The corresponding
behaviour is shown in figure \ref{fig:CA}, where for a wall attraction of
$\alpha_w \sigma^3 / \depthLJ=0.55$ \newB{corresponding to $\thYoung =
134.14^\circ$}, the numerical parameters $y_{\text{max}}$ and $\theta_n$ are varied. 
It is seen that for all values of $y_{\max}$ and $\theta_n$ the contact angle approaches
$\theta_{\text{Y}}$ for increasing $y$, before converging to $\theta =
\theta_n$ near $y=y_{\max}$ due to the imposed boundary condition.
For reference, the principal results presented in figure~\ref{fig:DensitySlices135} were computed on
a grid with $45 \times 75$ collocation points and parameters $y_{\text{max}} = 35\sigma$ and
$\theta_n = \{135^\circ,120^\circ,90^\circ,60^\circ,40^\circ\}$ for the different rows, respectively.

\section{Fluid structure in the vicinity of the contact line \label{sec:DensityStructure}}
\begin{figure}
	\centering
		\includegraphics[width=15cm]{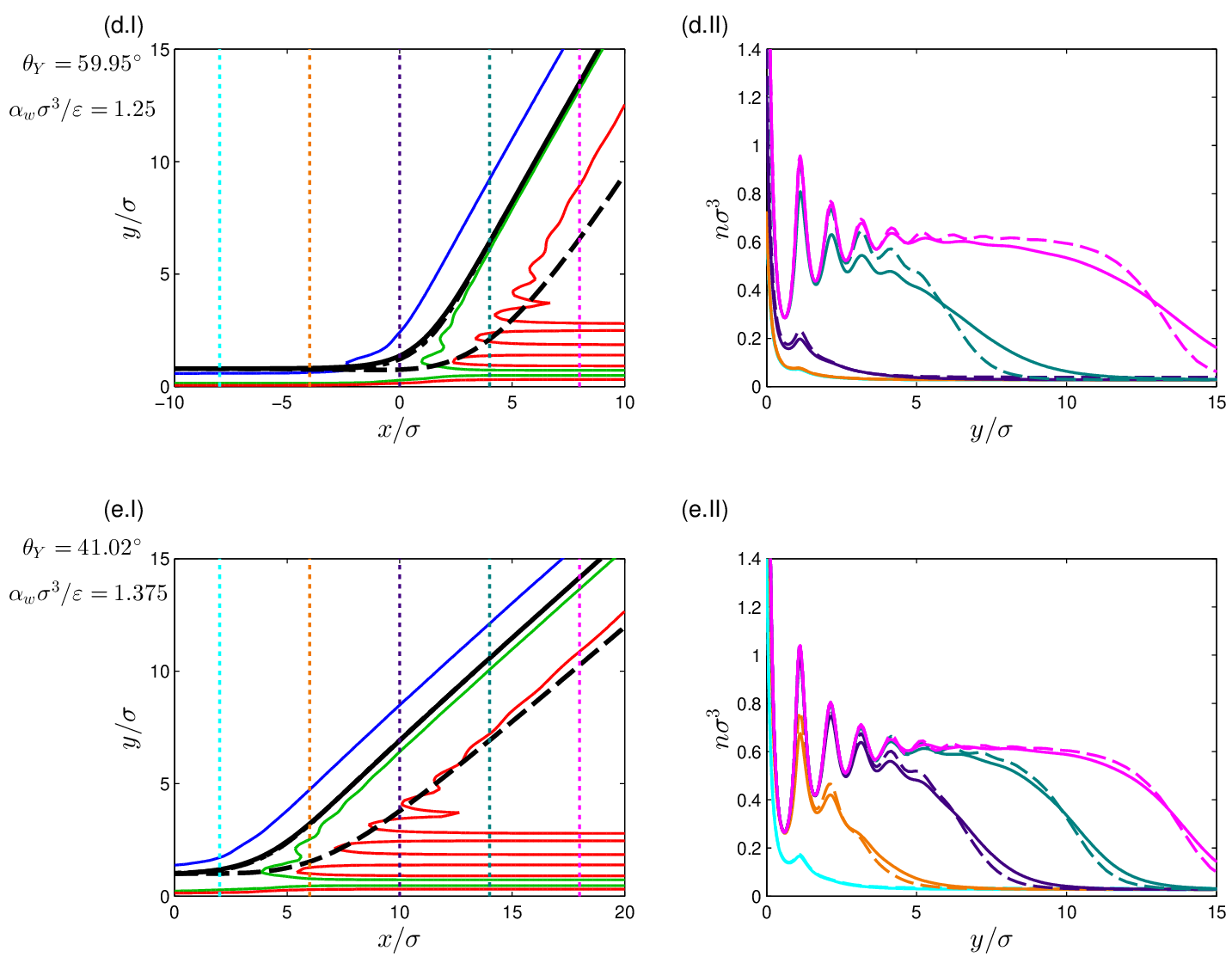}
	\caption{Contour lines for the number density (left column, \new{subfigures I}) and density profiles as \new{a} function of the distance to the
    substrate at various positions $x$ along the substrate (right column,
    \new{subfigures II}). 		
	In the left column, the contour lines correspond to number densities $\klamm{\nDensity - \nDensityV}/\klamm{\nDensityL - \nDensityV} = \{0.05,0.5,0.95\}$ from left to right.
The height profiles $h_{\text{I,II,III}}$, defined through equation (\ref{eq:MinimizingHamiltonian:Condition}) with boundary conditions (\ref{eq:h:BC1})-(\ref{eq:h:BC2}) and equation (\ref{eq:Def_HIII}),
are depicted by black dash-dotted, dashed and solid lines, respectively, $h_{\text{I}}$ being virtually indistinguishable from $h_{\text{III}}$.
		The solid lines in the right column represent the 2D density profile, plotted along the dashed vertical lines of corresponding colour in the left column figures.				
		These density profiles are compared to the equivalent planar off-saturation liquid or vapour film of the same adsorption film thickness, drawn with dashed lines.		
	\label{fig:DensitySlices135}
	}
\end{figure}

Figure \ref{fig:DensitySlices135} reveals the density structure for a fluid
in the vicinity of the contact line for different wall strengths. It can be
seen that depending on the wall strength parameter $\alpha_{\text{w}}$, the
contact density at the wall for the wall-liquid interface changes
significantly. In particular, we have checked the consistency of the observed
behaviour with the wall-fluid virial equation \cite{Herring:2010vn}
\begin{align}
p %&
= - \int_{-\infty}^\infty \nDensity(y) V'_{\text{ext}}(y) \dI y %\notag \\
%&
= \nDensity(0) - \int_{0}^\infty \nDensity(y) V'_{\text{ext}}(y) \dI y,
\end{align}
where $\nDensity(0)$ stems from the delta-function contribution to $V'_{\text{ext}}$ at $y=0$.

\begin{figure}
	\centering
		\includegraphics[width=15cm]{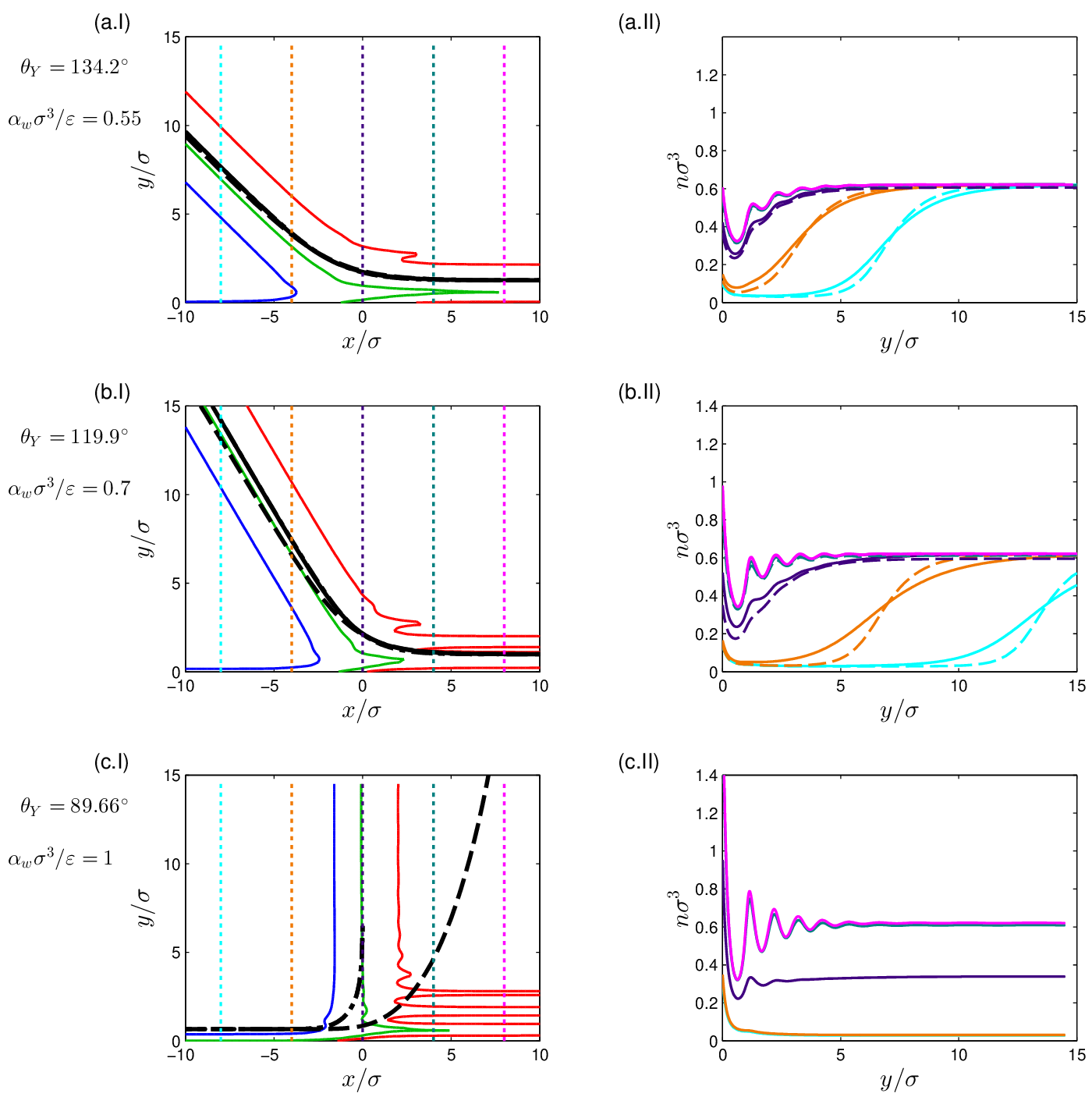}
	\caption*{For caption, see next page.}	
	\notag
\end{figure}

\begin{figure}[ht]
	\centering		
		\includegraphics[width=8cm]{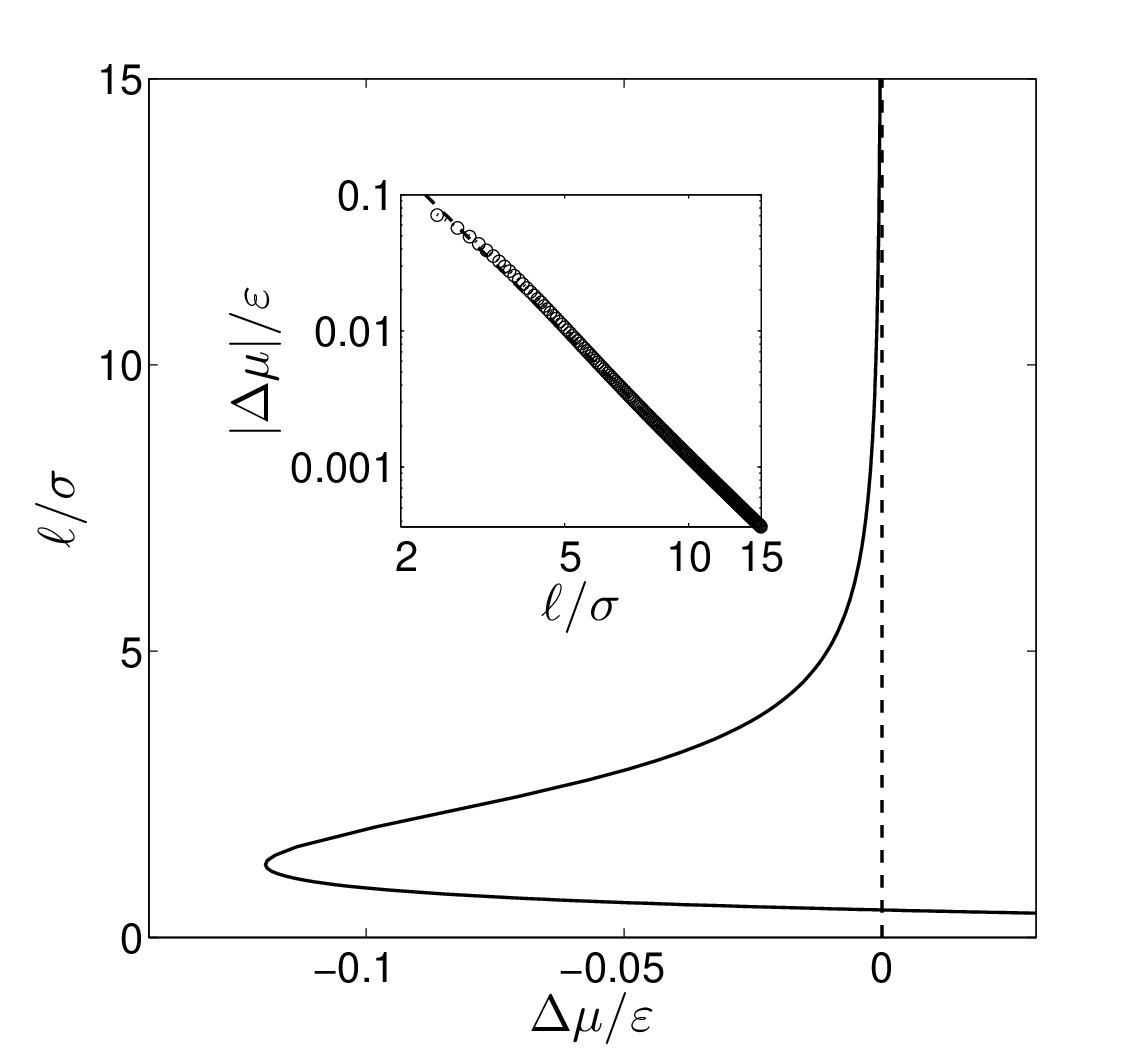}
	\caption{Plot of the adsorption isotherm for a dewetting scenario with wall attraction \new{of} $\alpha_{\text{w}} \LJdiam^3/\depthLJ = 0.7$,
    corresponding to $\theta_{\text{Y}} = 119.9^\circ$.
		The inset shows the asymptotic behaviour
	for \newB{large} $\filmThickness$, as $\Delta \mu \sim \filmThickness^{-3}$, \new{where the} dashed line is a fit
    for $\filmThickness \in [10 \LJdiam,15 \LJdiam]$ to $\Delta \mu = a
    \filmThickness^{-3}$, \new{with computed} coefficient $a = -1.21 \depthLJ \LJdiam^3$. \new{In the inset} individual
    DFT computations of the equilibrium density \new{are marked with circles and are connected by the solid line \newB{in the main plot} for clarity.}}		
	\label{fig:AdsorptionIsotherm41}
\end{figure}

The density plots at different positions in $x$ across the contact line in
the right column of figure \ref{fig:DensitySlices135} provide an insight as
to how the transition between a wall-vapour and a wall-liquid interface leads
to a quasi step-like increase of the density in the contour plots. We note
that this transition is accompanied by a gradual increase of the distance
between the liquid-vapour interface \newB{and} the wall. A similar transition
can be observed when gradually varying the chemical potential for a fluid
film in contact with a planar wall. A typical example \newB{of} the
bifurcation diagram, also widely denoted as the adsorption isotherm,
representing this transition, is shown in figure
\ref{fig:AdsorptionIsotherm41}, where the film thickness $\filmThickness$ of
the liquid or vapour film, defined by
\begin{align}
\filmThickness \defi&  \frac{1}{\Delta \nDensity} \int_0^\infty |\nDensity(\infty) - \nDensity(y)| \dI y  \label{eq:AdsorptionFilmThickness:Def}\\
\qquad\text{with} \qquad \Delta n =& \nDensityL  - \nDensityV,
\end{align}
is plotted versus the deviation of the chemical potential from its saturation
value $\Delta \chemPot$. In particular, figure \ref{fig:AdsorptionIsotherm41} shows the
behaviour for a dewetting scenario of a growing vapour film. Each point on the adsorption isotherm represents a density
profile which satisfies the Euler-Lagrange equation
(\ref{eq:EulerLagrangeEquation}) for a planar configuration.
As saturation is approached, the adsorption isotherm satisfies the expected inverse cubic decay of $\Delta \chemPot$ with $\filmThickness$ 
for systems with dispersion forces~\cite{DietrichNap:1991}, such as shown in the inset of figure \ref{fig:AdsorptionIsotherm41}.

These density profiles are compared in the right column of figure 
\ref{fig:DensitySlices135} with density profiles across the contact line
which have the same adsorption (\ref{eq:AdsorptionFilmThickness:Def}). We
note that the contact line is computed at saturation chemical potential,
whereas the chemical potential for the density profiles of the adsorption
isotherm is naturally off-saturation. Nevertheless, the result is unexpected
and shows a surprisingly good agreement, where for large film thicknesses,
the density profiles at the liquid-vapour interfaces differ because for a
contact line the liquid-vapour interface is at an angle to the wall, while
the dashed lines always describe planar films.

\section{Hamiltonian approaches, Derjaguin-Frumkin route and disjoining pressure \label{sec:HamiltonianApproaches}}

In a coarse-grained description of the contact line, the two-dimensional density profile is reduced to a height profile $h(x)$
representing the liquid-vapour interface~\cite{Lipowsky:1987,Mikheev:1991pi}. At equilibrium, this height profile minimizes the Hamiltonian \cite{Herring:2010vn}
\begin{align}
H[h] = \int_{-\infty}^\infty \klammCurl{ \surfaceTensionLV \klamm{ \sqrt{ 1+ (h')^2 } -1 } + V(h)  } \dI x, \label{eq:Hamiltonian_h}
\end{align}
where $h' = {\dI h}/{\dI x}$ is the slope of the interface and $V(h)$ is the effective interface potential.
The first term in (\ref{eq:Hamiltonian_h}) accounts for the excess energy stored through the surface tension due to the curvature of the liquid-vapour interface, while
the second term accounts for corrections to the Hamiltonian due to the presence of the substrate. This not only includes direct attractive forces between fluid and wall particles, but also
corrections due to the distorted fluid density profile caused by the presence of the wall.
The effective interface potential $V$ is linked to the disjoining pressure $\DisjoiningPressure$ by
\begin{align}
\DisjoiningPressure\klamm{h}  \defi - \frac{\dI V}{\dI h}. \label{eq:DisjoiningPressureVh}
\end{align}

Usually, (\ref{eq:Hamiltonian_h}) is only applied in the lubrication approximation. For larger slopes, both
the separate inclusion of the effective surface potential and surface energy~\cite{Pismen:2001fk}
as well as the functional dependence of $V$ on $h$ alone, as opposed to a functional dependence on $h(x)$, were put into question \cite{Henderson:2011:EPJST:DisjoiningPressure,MacDowell:2011:ResponseEPJST,Henderson:EPJST:ResponseMacDowell,Henderson:NoteContinuingContactLine}. Here, we test 
for different disjoining pressure definitions whether (\ref{eq:Hamiltonian_h}) may be used to to define height profiles for a large range of contact angles.

In \cite{Nold:2014:FluidStructure} we have compared height profiles resulting
from minimizing (\ref{eq:Hamiltonian_h}) with two different definitions of
the disjoining pressure for contact angles $\theta < 90^\circ$. We note that
these disjoining pressure definitions are different from phenomenological
analytical models such as used e.g. in \cite{Schwartz:1998,Sibley:JengMath:2014} in that they are
obtained directly from DFT computations, and therefore include the full
information of hard-sphere as well as the attractive particle interactions.
The first disjoining pressure definition we consider is based on the
celebrated Derjaguin and Frumkin theory \cite{Derjaguin:1936,Frumkin1938I}:
\begin{align}
\DisjoiningPressure_{\text{I}}\klamm{\filmThickness} \defi  - \Delta \chemPot \Delta \nDensity \times
\left\{
\begin{array}{ll}
1 & \text{for }  \nDensity|_{y=\infty} = \nDensityV\\
-1 & \text{for } \nDensity|_{y=\infty} = \nDensityL
\end{array}
 \right. , \label{eq:DisjoiningPressureAdsorptionIsotherm}
\end{align}
for a system at saturation chemical potential $\chemPot_{\text{sat}}$, and where
\begin{align}
\Delta \chemPot &= \chemPotN_{\text{eq}}\klamm{\filmThickness} - \chemPotN_{\text{sat}}.
\end{align}
$\chemPotN_{\text{eq}}$ is the chemical potential at which a film of thickness $\filmThickness$ is at equilibrium, such as depicted in the adsorption isotherm in figure \ref{fig:AdsorptionIsotherm41}.

The first case of (\ref{eq:DisjoiningPressureAdsorptionIsotherm}), \newB{$\nDensity|_{y=\infty} = \nDensityV$}, describes a
wetting scenario where the density at infinite distance from the wall
corresponds to the equilibrium vapour density. In this case, a liquid film
will slowly build as the chemical potential reaches its saturation value. In
contrast, the dewetting case \newB{$\nDensity|_{y=\infty} = \nDensityL$}
describes a vapour film in a bulk \new{liquid} environment, as studied in
figure \ref{fig:AdsorptionIsotherm41}. The sign switch in
(\ref{eq:DisjoiningPressureAdsorptionIsotherm}) originates from the sign
difference between the density in the film vs. the bulk density. We note that
contact lines with contact angle $\thYoung > 90^\circ$ are described by a
vapour film of varying height, whereas contact lines with contact angle
\newB{$\thYoung < 90^\circ$} are described by a liquid film of varying
height.

\new{As an alternative} to the Derjaguin and Frumkin definition of the disjoining pressure (\ref{eq:DisjoiningPressureAdsorptionIsotherm}), one can define the disjoining pressure based on the normal force balance at the substrate. The disjoining pressure is then defined as the excess pressure acting on the substrate due to the deviation from the equilibrium density profile, caused e.g.\ by the boundary conditions imposed on the system~\cite{Herring:2010vn,Henderson:2011:EPJST:DisjoiningPressure}
\begin{align}
\DisjoiningPressure_{\text{II}}\klamm{x} \defi - \int_{-\infty}^\infty  \klamm{ \nDensity(x,y) - \nDensity(\infty,y) } V_{\text{ext}}'(y) \dI y. \label{eq:SumRuleDisjoiningPressue2D}
\end{align}
Note that $\nDensity(x,y) V_{\text{ext}}'(y)$ is the force acting through the
external potential---representing the wall---on the fluid element at point
$(x,y)$. In our case, $\nDensity(x,y)$ is the density profile originating
from a 2D DFT computation of the contact line, and hence
$\DisjoiningPressure_{\text{II}}$ is a quantity containing information of the
full 2D equilibrium density profile; in contrast
(\ref{eq:DisjoiningPressureAdsorptionIsotherm}) is derived from planar 1D
computations.

The equilibrium height profiles $h_{\text{I}}$ and $h_{\text{II}}$
corresponding to the disjoining pressures $\DisjoiningPressure_{\text{I}}$
and $\DisjoiningPressure_{\text{II}}$, respectively, are obtained by
minimizing the Hamiltonian (\ref{eq:Hamiltonian_h}), leading to the defining
equation for $h_{\text{I/II}}$
\begin{align}
- \DisjoiningPressure_{\text{I}/\text{II}} = \surfaceTensionLV \frac{\dI}{\dI x} \klamm{ \frac{h_{\text{I}/\text{II}}'}{\sqrt{1+ (h_{\text{I}/\text{II}}')^2}} }, \label{eq:MinimizingHamiltonian:Condition}
\end{align}
with boundary conditions
\begin{align}
\lim_{x\to -\infty } h_{\text{I}} = h_0 \qquad   \text{for } \thYoung < 90^\circ, \label{eq:h:BC1}
\end{align}
and
\begin{align}
\lim_{x\to \infty } h_{\text{I}} = h_0 \qquad  \text{for } \thYoung > 90^\circ, \label{eq:h:BC2}
\end{align}
where $h_0$ is the film thickness representing the wall-vapour interface in
the wetting case and the wall-liquid interface in the drying case. We note
that $h_0$ corresponds to the (finite) value at $\Delta \chemPot = 0$ of the
adsorption isotherm in \new{figure \ref{fig:AdsorptionIsotherm41}}. Given
that $\DisjoiningPressure_{\text{I}}$ is a function of \new{$h$, and not of
$x$ directly}, (\ref{eq:MinimizingHamiltonian:Condition}) for $h_{\text{I}}$
is an autonomous ordinary differential equation. This means that with
(\ref{eq:h:BC1}),
\newB{(\ref{eq:h:BC2})} $h_{\text{I}}$ is translationally invariant in $x$.
For simplicity, in figures \ref{fig:DensitySlices135} and
\ref{fig:DisjoiningPressures}, we depict one representative plot for $h_{\text{I}}$ or
$\DisjoiningPressure_{\text{I}}(h_{\text{I}})$.

The ordinary differential equation (\ref{eq:MinimizingHamiltonian:Condition})
defining the film heights $h_{\text{I/II}}$ can also be interpreted as a form
of the Young-Laplace equation for a pressure jump across a fluid interface,
where the left hand side describes the difference between the pressure acting
on the substrate and the fluid pressure at $y=\infty$, while the right hand
side represents the product of the surface tension with the curvature of the
interface.

Integrating (\ref{eq:MinimizingHamiltonian:Condition}) with respect to $x$ and $h$, respectively, leads to the normal-force balance of Young's equation
\begin{align}
-\int_{-\infty}^\infty \DisjoiningPressure_{\text{I}/\text{II}}(x) \dI x = \surfaceTensionLV \sin \theta_{\text{Y},\text{I}/\text{II}}, \label{eq:SumRuleNormalForceBalance}
\end{align}
and the important expression of Derjaguin-Frumkin theory~\cite{derjaguin1986properties,Schwartz:1998}
\begin{align}
- \int_{h_0}^\infty \DisjoiningPressure_{\text{I}/\text{II}}\klamm{h} \dI h = \surfaceTensionLV \klamm{1 - |\cos \theta_{\text{Y},\text{I}/\text{II}}|}, \label{eq:FrumkinResultDisjoiningPressure1}
\end{align}
where $\theta_{\text{Y},\text{I}/\text{II}} \in [0,180^\circ]$ corresponds to the limiting slope of the height profiles $h_{\text{I,II}}$, respectively, at distances far away from the wall:
\begin{align}
\theta_{\text{Y},\text{I}/\text{II}} = \lim_{h_{\text{I/II}} \to \infty} \tan^{-1} \klamm{ h'_{\text{I/II}}(x)}. \label{eq:definethetaI_II}
\end{align}
Equation (\ref{eq:FrumkinResultDisjoiningPressure1}) can be interpreted as a
force balance in direction parallel to the substrate. For $\thYoung<
90^\circ$, the right hand side of the equation represents the forces of the
liquid-vapour interface acting in the negative $x$-direction. For $\thYoung >
90^\circ$, the height profile decreases from $\infty$ to $h_0$ as $x$ increases. Due to this
inversion of the height profile, (\ref{eq:FrumkinResultDisjoiningPressure1})
represents the force balance in the positive $x$-direction. The force of the
liquid-vapour interface acting in the positive $x$-direction is
$\surfaceTensionLV$, whereas the force acting in the negative direction is
$\surfaceTensionLV |\cos \thYoung|$. We note that here, the modulus accounts
for the fact that for $\thYoung > 90^\circ$, $\cos \thYoung < 0$, given that
we have defined $\theta_{\text{Y},\text{I}/\text{II}} \in [0,180^\circ]$, as
opposed to allowing for negative values of $\theta_{\text{Y,I/II}}$ in
(\ref{eq:definethetaI_II}).

Since both sum rules are derived from (\ref{eq:MinimizingHamiltonian:Condition}),
$\theta_{\text{Y},\text{I}/\text{II}}$ in equations
(\ref{eq:SumRuleNormalForceBalance}) and
(\ref{eq:FrumkinResultDisjoiningPressure1}) are equivalent and ultimately,
both height profiles converge to the slope dictated by the Young contact
angle. Thus $\theta_{\text{Y},\text{I}/\text{II}}$ both correspond to $\thYoung$
defined in the Young equation (\ref{YoungEquation}). We will exploit this
property to estimate the accuracy of our numerical method.

\begin{table}[ht]
	\centering
		\begin{tabular}{llllll}
		\toprule
			$\alpha_w \sigma^3/\varepsilon$ & $\theta_{\text{Y}}$  & $-\int_{h_0}^\infty \DisjoiningPressure_{\text{I}}\klamm{h} \dI h$ & $\theta_{\text{Y,I}}$ &
			$-\int_{-\infty}^\infty \DisjoiningPressure_{\text{II}}\klamm{x} \dI x$
			&$\theta_{\text{Y,II}}$\\\midrule
			$0.55$ & $134.2^\circ \pm 0.1^\circ$ & $0.103\pm 0.002$ & $134.5^\circ \pm 0.4^\circ$ & $-0.244 \pm 0.005$ & $135.2^\circ \pm 1.1^\circ$\\
			$0.7$ & $119.9^\circ\pm  0.05^\circ$ & $0.172\pm 0.003$ & $120.3^\circ \pm 0.5^\circ$ & $-0.298 \pm 0.002$ & $120.5^\circ \pm 0.7^\circ$ \\
			$1.0$ & $89.6^\circ \pm 0.1^\circ$ & $0.345 \pm 0.001$ & $89.8^\circ \pm 0.2^\circ$ & $-0.3463 \pm 10^{-4}$ & ($\star$) \\
			$1.25$ & $59.9^\circ \pm 0.1^\circ$ & $0.173 \pm 0.001$ & $60.0^\circ \pm 0.2^\circ$ &  $-0.297 \pm 0.003$ & $59.1^\circ \pm 0.8^\circ$\\
			$1.375$ & $41.0^\circ\pm 0.1^\circ$ & $0.085 \pm 0.001$ & $41.1^\circ \pm 0.2^\circ$ & $-0.234 \pm 0.007$ & $42.5^\circ \pm 1.6^\circ$\\\bottomrule
		\end{tabular}
		\caption{Comparison of $\theta_{\text{Y}}$ as defined in (\ref{YoungEquation}), the contact angles $\theta_{\text{I,II}}$ defined through (\ref{eq:definethetaI_II}) as well as the absolute errors of the integrals on the left hand sides of equations (\ref{eq:SumRuleNormalForceBalance}) and (\ref{eq:FrumkinResultDisjoiningPressure1}), respectively. ($\star$): Here, the integral expression gives $\sin \theta_{\text{Y,II}} = 1.0001 \pm 0.0001$, such that an estimate for $\theta_{\text{Y,II}}$ cannot formally be given.}
		\label{tab:thetaY_I_II_Comparison}
\end{table}

In table \ref{tab:thetaY_I_II_Comparison}, numerical values for the integrals
of the disjoining pressures are given. Error bounds $\Delta$ are estimated by
comparing the integral expressions with $\surfaceTensionLV \sin \thYoung$ and
$\surfaceTensionLV \klamm{1 - |\cos \thYoung|}$, respectively. These error
bounds are then used to estimate error bounds of $\theta_{\text{Y,I/II}}$ by
\begin{align}
\Delta \theta_{\text{Y,II}} = \left| \frac{\Delta \klammCurl{-\int_{-\infty}^\infty \DisjoiningPressure_{\text{II}}(x) \dI x}}{\surfaceTensionLV \cos \theta_{\text{Y},\text{II}}} \right|
\qquad \text{and} \qquad
\Delta \theta_{\text{Y,I}} = \left|\frac{\Delta \klammCurl{- \int_{h_0}^\infty \DisjoiningPressure_{\text{I}}\klamm{h} \dI h}}{ \surfaceTensionLV \sin \theta_{\text{Y},\text{I}} }\right|.
\end{align}
The above formulations can be derived from
(\ref{eq:SumRuleNormalForceBalance}) and
(\ref{eq:FrumkinResultDisjoiningPressure1}) by using
$\theta_{\text{Y},\text{I/II}} + \Delta \theta_{\text{Y},\text{I/II}}$  and
linearly expanding to first order in $\Delta \theta_{\text{Y},\text{I/II}}$
the right hand side of the respective equation around
$\theta_{\text{Y},\text{I/II}}$. Finally, we compare the film height profiles
$h_{\text{I}}$ and $h_{\text{II}}$ with the adsorption film thickness
\begin{align}
h_{\text{III}}(x) \defi \frac{1}{\Delta \nDensity} \int_0^\infty |\klamm{\nDensity(x,y) - \nDensity(x,\infty)}| \dI y, \label{eq:Def_HIII}
\end{align}
\new{which is the 2D generalisation of (\ref{eq:AdsorptionFilmThickness:Def}).}
This allows us to define a disjoining pressure suggested by the adsorption film height, obtained by inserting $h_{\text{III}}$ into (\ref{eq:MinimizingHamiltonian:Condition}), giving the rescaled curvature
\begin{align}
- \DisjoiningPressure_{\text{III}}(h) \defi \surfaceTensionLV \frac{\dI}{\dI x} \klamm{ \frac{h_{\text{III}}'}{\sqrt{1+ (h_{\text{III}}')^2}} }.
\end{align}
In figure~\ref{fig:DensitySlices135}, we compare the height profiles $h_{\text{I}-\text{III}}$ in the vicinity of the contact line for a wide range of wall attractions.

\section{Discussion and conclusion \label{sec:Conclusion}}

We have scrutinized the fluid structure and its properties in the vicinity of
a three-phase contact line by employing a DFT-FMT model. In particular, we
presented density profiles slice by slice as we sweep through the contact
line region and we contrast the density profiles with the profile of a planar
liquid film on a substrate, but with the same film thickness, demonstrating
that the two are quite similar. We also scrutinized the ability of
Derjaguin-Frumkin theory~\cite{Derjaguin:1987:50YearsOfSurfaFceScience} for
planar liquid films on a substrate to predict the height profile at the
contact line and we offered a unified Derjaguin-Frumkin treatment of the
contact line for $\thYoung < 90^\circ$ and $\thYoung > 90^\circ$ by
appropriately extending the boundary conditions for the disjoining pressure
equation to account for the case $\thYoung > 90^\circ$.

\begin{figure}[ht]
	\centering
		\includegraphics[width=14cm]{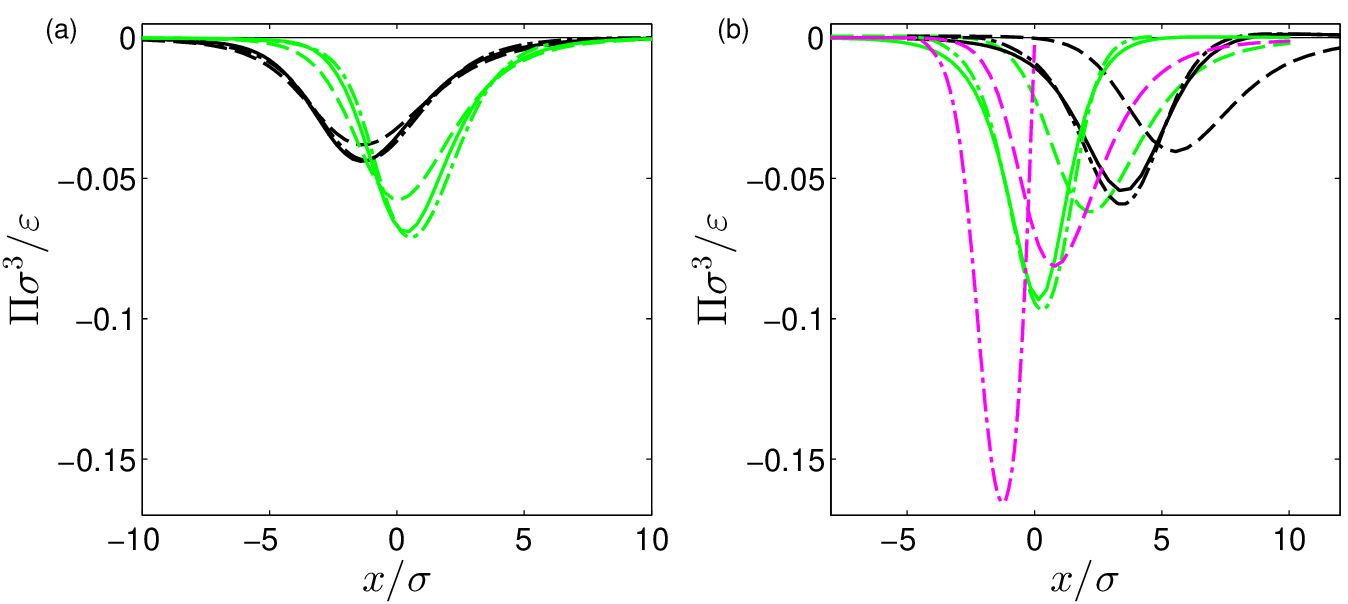}
		\caption{Plots of different disjoining pressure definitions for different wall attractions. \new{Dash-dotted, dashed and solid} lines depict disjoining pressures
        \new{$\DisjoiningPressure_{\text{I}}$, $\DisjoiningPressure_{\text{II}}$ and $\DisjoiningPressure_{\text{III}}$}, respectively.
        In \new{subfigure (a)}, the black and green lines show data for $\alpha_w \sigma^3/\varepsilon = 0.55$ and $0.7$, respectively, \new{whilst in (b)},
        the black, green and magenta lines show data for $\alpha_w \sigma^3/\varepsilon = 1.375, 1.25$ and $1.0$, respectively.}
			\label{fig:DisjoiningPressures}
\end{figure}

In figure \ref{fig:DensitySlices135} we plot the height profiles
$h_{\text{I/II/III}}$ for contact angles in the region $40^\circ < \thYoung <
135^\circ$ and compare them with the contour lines of the density. The figure summarizes
some of the main results of our study as far as the behaviour close to the
contact line is concerned. Additional information on this can be extracted
from figure \ref{fig:DisjoiningPressures} where we compare the disjoining
pressure profiles $\DisjoiningPressure_{\text{I/II/III}}$. An observation we
made in our previous study in~\cite{Nold:2014:FluidStructure} for contact
angles $\thYoung <90^\circ$, was that the location of maximal curvature for
the height profile $h_{\text{II}}$ is shifted towards the fluid phase if
compared with the adsorption height profile $h_{\text{III}}$. This
observation can also be made in figures \ref{fig:DensitySlices135}
\new{(g,i)} and in figure \ref{fig:DisjoiningPressures} \new{(b)}. However,
this does not occur to the same extent in cases where $\thYoung >90^\circ$---such as
observed in figures \ref{fig:DensitySlices135} (a,c) and in figure
\ref{fig:DisjoiningPressures}
\new{(a)}.

Furthermore, the maximal absolute curvature of the height profile
$h_{\text{II}}$ \newB{(see dashed lines in figures \ref{fig:DensitySlices135}
and \ref{fig:DisjoiningPressures})} is lower than the maximal absolute
curvature of the adsorption film height $h_{\text{III}}$ \newB{(see solid
lines in figures \ref{fig:DensitySlices135} and
\ref{fig:DisjoiningPressures})}. This can best be seen in figure
\ref{fig:DisjoiningPressures} \newB{(we note that the disjoining pressure
corresponds to the rescaled curvature of the corresponding height profile)}.
While the difference is less pronounced for large contact angles $\thYoung >
90^\circ$, it is still observable. In contrast, the film thickness
$h_{\text{I}}$ \newB{(see dash-dotted lines in figures
\ref{fig:DensitySlices135} and \ref{fig:DisjoiningPressures})} based on the
adsorption isotherm, agrees very well with $h_{\text{III}}$, often to the
point of being virtually indistinguishable (compare the left column of figure
\ref{fig:DensitySlices135}).

For a varying height profile, here enforced by the
boundary conditions, we have studied two conflicting definitions of the disjoining
pressure---one based on the adsorption isotherm, the other based on the
normal force balance. These two definitions lead to distinct height profiles,
which suggest that the use of the disjoining pressure based on the adsorption
isotherm is more appropriate, given the good agreement of the corresponding
height profile with the adsorption height profile. This is \new{somewhat}
surprising, given that the disjoining pressure based on the normal force
balance $\DisjoiningPressure_{\text{II}}$ contains information from the full
equilibrium 2D density profile, whereas $\DisjoiningPressure_{\text{I}}$ is
derived from purely 1D computations.

At the same time the behaviour of $\DisjoiningPressure_{\text{II}}$ is such
that the maximum absolute normal pressure acting on the substrate is lower
than the curvature of the adsorption height profile would suggest. Also, for
$\thYoung < 90^\circ$, the maximal normal pressure does {\it not} act in the
vicinity of the contact line, but instead at a slightly shifted position
towards the liquid phase. This interpretation could be of interest
for the nanoscale behaviour of contact lines at soft substrates, such as
considered e.g.\ by Lubbers \etal~\cite{DubbersSnoeijer:2014:SoftSolids}.

The special case of $\thYoung$ being very close to $90^\circ$, such as
depicted in figure \ref{fig:DensitySlices135} \new{(e,f)} for $\alpha_w
\sigma^3/\varepsilon = 1.0$, as well as the magenta lines in figure
\ref{fig:DisjoiningPressures} \new{(b)}, deserves a comment. In this case,
the density at very large distances from the wall $\nDensity|_{y \to \infty}$
depends on the position $x$, and hence does not allow for the definition of
an adsorption height profile $h_{\text{III}}$ through (\ref{eq:Def_HIII}).
While the disjoining pressure $\DisjoiningPressure_{\text{I}}$ based on the
adsorption isotherm has a very high absolute maximum, the absolute maximum of
$\DisjoiningPressure_{\text{II}}$ is less pronounced. Also, \new{the} width
of $\DisjoiningPressure_{\text{II}}$ corresponds roughly to the width of the
interface and is slightly shifted towards the fluid phase.

An important observation, therefore, is that the maximal normal pressure
acting on the substrate does not correspond with the maximal curvature of the
adsorption film thickness or the maximal value of the Derjaguin-Frumkin
disjoining pressure $\DisjoiningPressure_{\text{I}}$. One reason for the
softening of the normal pressure profile could be the width of the fluid
interface. In particular, one can observe in figure
\ref{fig:DisjoiningPressures} (b), that the width of
$\DisjoiningPressure_{\text{II}}$ for $\thYoung \approx 90^\circ$, denoted by
the dashed magenta line, corresponds approximately with the width of the
liquid-vapour interface.

It is noteworthy that the main limitation of the model is that its mean-field
nature does not include the description of thermal
fluctuations~\cite{ArcherEvans:2013,EvansHendersonHoyle:1993,MacDowellBenet:2014}.
Inclusion of thermal fluctuations, which become more pronounced with increasing 
film thicknesses $\filmThickness$, leads to a broadening of the liquid-vapour interface
and a renormalization of the dependence of $\filmThickness$ on the chemical potential deviation from saturation $\Delta \chemPot$~\cite{ArcherEvans:2013} is needed.
A detailed recent study based on molecular simulations and experiments
has found that thermal fluctuations lead to an effective film-height dependent surface tension $\surfaceTensionLV(\filmThickness)$ in (\ref{eq:Hamiltonian_h})~\cite{MacDowellBenet:2014}.
A final conclusion about the effect on thermal fluctuations for the results presented here could be reached by a molecular simulations study in the spirit of Herring and Henderson's analysis~\cite{Herring:2010vn},
but including dispersion forces and a comparison with the corresponding Derjaguin-Frumkin disjoining pressure. This, however, is beyond the scope of the present study.

The important observation made here is that in a mean-field model, disjoining pressures
obtained from planar films via the Derjaguin-Frumkin route do allow us to
predict with good accuracy the structure of the contact line, hence implying
a negligeble contribution of non-locality. It would be interesting to see if this
holds for other settings, e.g. spherical droplets.

Of particular interest would also be to investigate very large contact angles
close to $180^\circ$, given interesting recent results in this
case~\cite{benilov2013contact} as well as the influence of surface roughness
and chemical heterogeneities which are known to influence wetting phenomena
substantially
(e.g.~\cite{PhysFluids_21_2009,Savva2011,PhysRevLett_104_2010,Raj2011}). We
shall address these and related issues in future studies.

\vspace*{0.5cm}

\section*{Acknowledgements}
We acknowledge financial support from ERC Advanced Grant No. 247031 and Imperial College through a DTG International Studentship.

%\bibliographystyle{plain}
%\bibliography{Bibliography}

\begin{thebibliography}{10}

\bibitem{ArcherEvans:2013}
A.~J. Archer, R.~Evans.
\textit{Relationship between local molecular field theory and density
  functional theory for non-uniform liquids.}
J. Chem. Phys., 138 (1) (2013), 014502.

\bibitem{Barker:1967rq}
J.~A. Barker, D.~Henderson.
\textit{Perturbation theory and equation of state for fluids. {II}. {A}
  successful theory of liquids.}
J. Chem. Phys., 47 (11) (1967), 4714--4721.

\bibitem{benilov2013contact}
E.~S. Benilov, M.~Vynnycky.
\textit{ Contact lines with a $180^\circ$ contact angle.}
J. Fluid Mech., 718 (2013), 481--506.

\bibitem{Bonn.20090527}
D.~Bonn, J.~Eggers, J.~Indekeu, J.~Meunier, E.~Rolley.
\textit{ Wetting and spreading.}
Rev. Mod. Phys., 81 (2009), 739--805.

\bibitem{Derjaguin:1987:50YearsOfSurfaFceScience}
B.~V. Derjaguin.
\textit{Some results from 50 years' research on surface forces.}
In {\em Surface Forces and Surfactant Systems}, volume~74 of {\em
  Progress in Colloid \& Polymer Science}, Steinkopff, (1987), 17--30. 
 
\bibitem{derjaguin1986properties}
B.~V. Derjaguin, N.~V. Churaev.
\textit{ Properties of water layers adjacent to interfaces.}
In Clive~A. Croxton, editor, Fluid interfacial phenomena, Wiley, New York, (1986), 663--738.

\bibitem{Derjaguin:1936}
B.~V. Derjaguin, E.~Obuchov.
\textit{ Anomalien d\"unner {F}l\"ussigkeitsschichten. {III.}}
Acta Physicochim. URSS, 5 (1) (1936), 1-22.

\bibitem{DietrichNap:1991}
S.~Dietrich, M.~Napi\'orkowski.
\textit{ {Analytic results for wetting transitions in the presence of van der
  Waals tails}.
Phys. Rev. A}, 43 (1991), 1861--1885.

\bibitem{Evans}
R.~Evans.
\textit{ The nature of the liquid-vapour interface and other topics in the
  statistical mechanics of non-uniform, classical fluids.}
Adv. Phys., 28 (2) (1979), 143--200.

\bibitem{EvansHendersonHoyle:1993}
R.~Evans, J.~R. Henderson, D.~C. Hoyle, A.~O. Parry, Z.~A. Sabeur.
\textit{ Asymptotic decay of liquid structure: oscillatory liquid-vapour
  density profiles and the Fisher-Widom line}.
Mol. Phys., 80 (4) (1993), 755--775.

\bibitem{Frumkin1938I}
A.~N. Frumkin.
\textit{\"Uber die Erscheinungen der Benetzung und des Anhaftens von
  Bl\"aschen. I}.
Acta Physicochim. URSS, 9 (313), (1938).

\bibitem{Getta:1998ly}
T.~Getta, S.~Dietrich.
\textit{Line tension between fluid phases and a substrate.}
Phys. Rev. E, 57 (1) (1998), 655--671.

\bibitem{Henderson:NoteContinuingContactLine}
J.~R. Henderson.
\textit{Discussion notes: Note continuing the discussion on the contact line
  problem.}
Eur. Phys. J. Special Topics, 197 (1) (2011), 129--130.

\bibitem{Henderson:EPJST:ResponseMacDowell}
J.~R. Henderson.
\textit{Discussion notes on ``{C}omputer simulation of interface potentials:
  {T}owards a first principle description of complex interfaces?'', by {L. G.}
  {M}ac{D}owell.}
Eur. Phys. J. Special Topics, 197 (1) (2011), 147--148.

\bibitem{Henderson:2011:EPJST:DisjoiningPressure}
J.~R. Henderson.
\textit{ Disjoining pressure of planar adsorbed films.}
Eur. Phys. J. Special Topics, 197 (1) (2011), 115--124.

\bibitem{Herring:2010vn} 
A.~R. Herring, J.~R. Henderson.
\textit{ Simulation study of the disjoining pressure profile through a
  three-phase contact line.}
J. Chem. Phys., 132 (8) (2010) 084702.

\bibitem{Lipowsky:1987}
R.~Lipowsky, M.~E. Fisher.
\textit{ Scaling regimes and functional renormalization for wetting
  transitions.}
Phys. Rev. B, 36 (4) (1987) 2126--2141.

\bibitem{DubbersSnoeijer:2014:SoftSolids}
L.~A. Lubbers, J.~H. Weijs, L.~Botto, S.~Das, B.~Andreotti, J.~H. Snoeijer.
\textit{ Drops on soft solids: free energy and double transition of contact
  angles.}
 J. Fluid Mech., (2014), 747, R1.

\bibitem{MacDowell:2011:ResponseEPJST}
L.~G. MacDowell.
\textit{ Discussion notes on ``{D}isjoining pressure of planar adsorbed
  films'', by {J}.{R}. {H}enderson.}
Eur. Phys. J. Special Topics, 197 (1) (2011),149--150.

\bibitem{MacDowellBenet:2014}
L.~G. MacDowell, J.~Benet, N.~A. Katcho, J.~M.~G. Palanco.
\textit{ Disjoining pressure and the film-height-dependent surface tension of
  thin liquid films: New insight from capillary wave fluctuations.}
 Adv. Colloid Interface Sci., 206 (2014), 150--171.

\bibitem{Malijevsky:2013:CriticalPointWedgeFilling}
A.~Malijevsk\'y, A.~O. Parry.
\textit{ Critical point wedge filling.}
Phys. Rev. Lett., 110 (16) (2012), 166101.

\bibitem{Merath:2008}
R.-J.~C. Merath.
\textit{ Microscopic calculation of line tensions}.
PhD thesis, Universit\"at Stuttgart, 2008.

\bibitem{Merchant:1992kx}
G.~J. Merchant, J.~B . Keller.
\textit{ Contact angles}.
Phys. Fluids A, 4 (3) (1992), 477--485.

\bibitem{Mermin:1965lo}
N.~D. Mermin.
\textit{Thermal properties of the inhomogeneous electron gas.}
Phys. Rev., 137 (5A) (1965), A1441--A1443.

\bibitem{Mikheev:1991pi}
L.~V. Mikheev, J.~D. Weeks.
\textit{Sum rules for interface {H}amiltonians.}
 Physica A, 177 (1) (1991), 495--504.

\bibitem{Nold:2014:FluidStructure}
A.~Nold, D.~N. Sibley, B.~D. Goddard, S.~Kalliadasis.
\textit{Fluid structure in the immediate vicinity of an equilibrium
  three-phase contact line and assessment of disjoining pressure models using
  density functional theory.}
Phys. Fluids, 26 (7) (2014) 072001.

\bibitem{Antonio2010}
A.~Pereira, S.~Kalliadasis.
\textit{Equilibrium gas-liquid-solid contact angle from density-functional
  theory.}
J. Fluid Mech., 692 (2012), 53--77.

\bibitem{Pismen:2001fk}
L.~M. Pismen.
\textit{ Nonlocal diffuse interface theory of thin films and the moving
  contact line.}
 Phys. Rev. E, 64 (2) (2001), 021603.

\bibitem{Rosenfeld:1989qc}
Y.~Rosenfeld.
\textit{Free-energy model for the inhomogeneous hard-sphere fluid mixture and
  density-functional theory of freezing.}
Phys. Rev. Lett., 63 (9) (1989), 980--983.

\bibitem{Roth:2010fk}
R.~Roth.
\textit{Fundamental measure theory for hard-sphere mixtures: a review.}
J. Phys.: Condens. Matter, 22 (6) (2010) 063102.

\bibitem{PhysFluids_21_2009}
N.~Savva, S.~Kalliadasis.
\textit{Two-dimensional droplet spreading over topographical substrates.}
Phys. Fluids, 21 (9) (2009), 092102.

\bibitem{Savva2011}
N.~Savva, S.~Kalliadasis.
\textit{Dynamics of moving contact lines: A comparison between slip and
  precursor film models.}
Europhys. Lett., 94 (6) (2011), 64004.

\bibitem{PhysRevLett_104_2010}
N.~Savva, S.~Kalliadasis, G.~A. Pavliotis.
\textit{Two-dimensional droplet spreading over random topographical
  substrates.}
Phys. Rev. Lett., 104 (8) (2010), 084501.

\bibitem{Schwartz:1998}
L.~W. Schwartz.
\textit{Hysteretic effects in droplet motions on heterogeneous substrates:
  Direct numerical simulation.}
Langmuir, 14 (12) (1998), 3440--3453.

\bibitem{Sibley:JengMath:2014}
D.~N. Sibley, A.~Nold, N.~Savva, S.~Kalliadasis.
\textit{A comparison of slip, disjoining pressure, and interface formation
  models for contact line motion through asymptotic analysis of thin two-dimensional droplet spreading.}
J. Eng. Math., {DOI}: 10.1007/s10665-014-9702-9, 2014.

\bibitem{Snoeijer:2008fk}
J.~H. Snoeijer, B.~Andreotti.
\textit{A microscopic view on contact angle selection.}
Phys. Fluids, 20 (5) (2008), 057101.

\bibitem{SnoeijerAndreotti:2013}
J.~H. Snoeijer, B.~Andreotti.
\textit{Moving contact lines: Scales, regimes, and dynamical transitions.}
Annu. Rev. Fluid Mech., 45 (2013), 269--292.

\bibitem{Trefethen_2000}
N.~L. Trefethen.
\textit{Spectral Methods in {MATLAB}}.
Vol. 10, SIAM, Philadelphia, 2000.

\bibitem{Raj2011}
R.~Vellingiri, N.~Savva, S.~Kalliadasis.
\textit{Droplet spreading on chemically heterogeneous substrates.}
Phys. Rev. E, 84 (3) (2011), 036305.

\bibitem{Wu-DFT}
J.~Wu.
\textit{Density functional theory for chemical engineering: {F}rom
  capillarity to soft materials.}
AIChE J., 52 (3) (2006), 1169--1193.

\bibitem{Peter2012}
P.~Yatsyshin, N.~Savva, S.~Kalliadasis.
\textit{Spectral methods for the equations of classical density-functional
  theory: {R}elaxational dynamics of microscopic films.}
J. Chem. Phys., 136 (12) (2012), 124113.

\bibitem{PeterPRE}
P.~Yatsyshin, N.~Savva, S.~Kalliadasis.
\textit{Geometry-induced phase transition in fluids: {C}apillary prewetting.}
Phys. Rev. E, 87 (2) (2013), 020402(R).

\end{thebibliography}

\end{document}